\documentstyle[twocolumn,prl,aps,epsfig]{revtex}

\begin{document}
\renewcommand{\thefootnote}{\fnsymbol{footnote}}
\sloppy

\newcommand \be{\begin{equation}}
\newcommand \bea{\begin{eqnarray} \nonumber }
\newcommand \ee{\end{equation}}
\newcommand \eea{\end{eqnarray}}
\newcommand{\rar}{\rightarrow}
\newcommand{\eq}{equation}
\newcommand{\eqs}{earthquakes}
\newcommand{\rp}{\right)}
\newcommand{\lp}{\left(}

\twocolumn[\hsize\textwidth\columnwidth\hsize\csname
@twocolumnfalse\endcsname

\title{Novel Mechanism for Discrete Scale Invariance in Sandpile Models}

\author{Matt Lee$^1$ and Didier Sornette$^{1-3}$ \\
$^1$ Institute of Geophysics and
Planetary Physics\\ University of California, Los Angeles, California 90095\\
$^2$ Department of Earth and Space Science\\
University of California, Los Angeles, California 90095\\
$^3$ LPMC, CNRS UMR6622 and Universit\'{e} de Nice-Sophia Antipolis\\ B.P.
71, Parc
Valrose, 06108 Nice Cedex 2, France}

\date{\today}
\maketitle

\begin{abstract}

Numerical simulations and a mean-field analysis of a sandpile model of
earthquake aftershocks
in 1d, 2d and 3d euclidean lattices
determine that the average stress decays in a punctuated fashion after a
main shock,
with events occurring
at characteristic times increasing as a geometrical series with a well-defined
multiplicative factor which is a function of the stress corrosion exponent,
the stress drop ratio and
the degree of dissipation. These results are independent of the discrete
nature of the lattice
and stem from the interplay between the threshold dynamics and the power
law stress
relaxation.

\end{abstract}

\vspace{5mm}
]

\narrowtext

Discrete scale invariance (DSI) \cite{Revue} is the partial breaking of
continuous
scale invariance \cite{Dubrulle} in which a system or an observable is
invariant only under
scaling ratios that are integer powers of a fundamental factor $\lambda$.
DSI leads to
complex critical exponents (or dimensions), i.e. to log-periodic corrections to
scaling, which reflect the existence of a discrete self-similar spectrum of
characteristic scales decorating the usual scale-free power law behavior.

Several mechanisms responsible
for this partial breaking of the continuous scale symmetry have been
expounded, which include build-in pre-existing hierarchy \cite{DIL},
intermittent
diffusion in discrete euclidean lattices \cite{BS} and cascades of
ultra-violet instabilities in
growth processes and rupture \cite{DLA}. Other situations are less well
understood but can be traced back to special technical properties such as
the non-unitary
structure of the underlying field theory describing the coarse-grained
properties of
animals \cite{Salsor} and of quenched disordered spin systems with
long-range interactions
\cite{Aharony}. Another example is
the gravitational collapse leading to critical black hole formation
described by a
a system of partial differential equations possessing an asymptotic solution
which can be understood from a renormalization group with a
limit cycle having a single unstable mode \cite{Blackhole}.

Here, we present a novel scenario for DSI based on the interplay between
the threshold dynamics characteristic of sandpile models and
a scale-free relaxation process \cite{Matt}. Specifically, we study a
conceptual sandpile model \cite{sandpile} of earthquake aftershocks on
a euclidean discrete $d$-dimensional cubic lattice with $L^d$ cells and
periodic boundary conditions with $d=1,2,3$. Each cell represents
a region which is
unloaded when an elementary fault is activated. We neglect the
tensorial nature of the stress field and consider an
anti-plane driving configuration in which loading and rupture
are controlled by a single shear stress component $V(\overrightarrow{x})$.

There are two distinct temporal phases. First,
the stress is uniformly
increased at a very slow rate on all cells to mimic the tectonic loading.
Due to the rupture and loading rules described below, the system
self-organises into a statistical stationary state, characterised by a power
law distribution of event sizes \cite{sandpile,Matt}.
Once this statistical stationarity state is established, we
freeze the loading
and the aftershock sequence starts, mimicking the aftermath of a great
earthquake. The second phase is characterised by the fact that the
aftershocks are not driven by the tectonic loading but by relaxation processes
as described below.

An initial stress threshold $B(\overrightarrow{x})$ is assigned to each cell
from a random uniform distribution in the interval $B_0[1-r,1+r]$. We
find similar results both for the
annealed and quenched version of the model, in which
either the thresholds are fixed or are resampled in the interval after
each rupture.
When the stress $V(\overrightarrow{x})$ in a cell at
$\overrightarrow{x}$ becomes larger or equal to
$B(\overrightarrow{x})$, the stress is re-distributed according to the rules
\begin{eqnarray} \label{hfql}
V(\overrightarrow{x})|^{\rm after} & = & V(\overrightarrow{x})|^{\rm
before} ~(1-\gamma)~,\\
V(\overrightarrow{x})|_{\rm nn}^{\rm after} & = &
V(\overrightarrow{x})|_{\rm nn}^{\rm before}
+ V(\overrightarrow{x})|^{\rm before}~ \frac{(1-\beta)\gamma }{2d}~.
\label{hfjqkq}
\end{eqnarray}
Rule (\ref{hfjqkq}) applies to each of the $2d$ nearest neighbours (n.n.)
of $\overrightarrow{x}$ carrying an initial stress
$V(\overrightarrow{x})|_{\rm nn}^{\rm before}$ which
evolves into $V(\overrightarrow{x})|_{\rm nn}^{\rm after}$.
Because the toppling criterion (\ref{hfql}) depends only on $V$ and
not on its gradient, the order of site toppling commute \cite{Dhar}.

$\gamma$ is the relative stress drop, with
$\gamma=1$ corresponding to a complete stress drop. $\beta$ is the
dissipation where
$1-\beta$ quantifies the amount of stress drop transfered to n.n. and is known
as the seismic efficiency.
For $\gamma=1$, (\ref{hfql}) and (\ref{hfjqkq}) are identical to the rules
used
in the non-conservative sandpile model \cite{Olami}, motivated from the
coupling of blocks to a rigid upper driving plate in the
Burridge-Knopoff model. Here, the dissipation accounts for
the loss of stress and of stored elastic energy due to an earthquake under
constant displacement conditions at the boundaries \cite{SOCHuang}.

In the second relaxation phase, the loading stops
and the thresholds decay in time according to the law
\begin{equation}
\label{af:ca-decay-law}
B\left( \overrightarrow{x},t\right) =B\left( \overrightarrow{x},t_{0}\right) -
\frac{[V\left( \overrightarrow{x},t\right)]^{\alpha }}{B\left(
\overrightarrow{x},t_{0}\right)}
~\left( t-t_{0}\right) ~.
\end{equation}
This model incorporates the mechanism of sub-critical crack growth and
stress corrosion \cite{Charles}, which has been proposed
as a possible delay mechanism for aftershocks \cite{Mecaafter,das81,Matt}. In
absence of loading,
events are triggered each time the
thresholds decay below the local stresses. When this occurs,
the stress redistribution obeys (\ref{hfql}) and (\ref{hfjqkq}).
The system eventually relaxes after a infinite time and in an intermittent
manner to an
equilibrium of zero stress on all elements of the lattice.
A closed conservative spring-block system has also been found to relax
in a infinite time
and in an intermittent manner to the zero-stress equilibrium with
self-organized
critical behavior \cite{Leung}.
It is this complex relaxation
that we study. It occurs via the triggering of
what can be called aftershocks which exhibit remarkable properties.
The results presented below are also found for a version of the model with
continuous
elasticity \cite{Matt,LeeKno} derived from Ref.\cite{Cowie}
and are probably robust features of the general
interplay between threshold dynamics and relaxation phenomena.

We show here the simulations for 2d systems of $20 \times 20$ elements
(simulations
have been performed with size up to $100 \times 100$ with no change of
results) and
in the annealed case where, after each toppling, thresholds are reassigned
from the uniform
distribution $B_0[1-r,1+r]$ with $r=0.75$. The system is up-dated by finding
the site closest to rupture and incrementing time, so that this site reaches
its threshold. Once a site becomes unstable due to either loading
(in the first phase) or by the decay of the threshold (in the second
phase), stress is
distributed to n.n. according to (\ref{hfql}) and (\ref{hfjqkq}).
The n.n. may also become unstable, releasing their
stress and the process continues until no further nodes are unstable, thus
defining an event. During the rupture process, time is ``freezed'' to ensure
a separation of time scales between rupture (fast) and loading/decay
(slow). When no further sites are unstable, loading or decay is continued
until the next event, when time ``freezes'' again. Typically $3 \cdot 10^3$ to
 $15 \cdot 10^3$ events were sampled in the decay regime. The results are
robust with respect to
heterogeneity level $r$, open or closed boundary conditions, and to the size
and dimension $d=1,2,3$ of the lattice.

\begin{figure}
\begin{center}
\epsfig{file=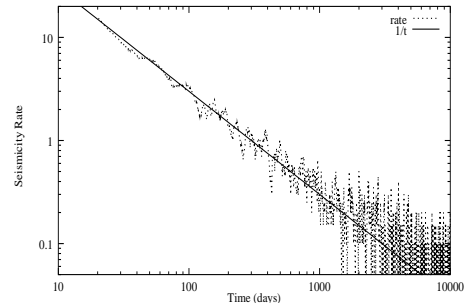, width=6cm, height=4cm}
\caption{\protect\label{afshrate} Power law decay of the rate of
aftershocks for a lattice of size $20 \times 20$ with
 $\alpha = 1$, $\gamma=1$ and $\beta = 0.5$. The solid line is $t^{-1}$.}
\end{center}
\end{figure}

Figure \ref{afshrate} shows the rate of aftershocks $n\lp t\rp$ as
a function of time after the loading has ceased. The $1/t$ decay is in
full agreement with Omori's law for real aftershock sequences \cite{Omori}
and is very robust over a wide range of parameters, i.e. $\alpha > 0$,
$\gamma >0$
and $\beta < 1$. Typical values for the earth are $10 \leq \alpha \leq 100$,
$\gamma \approx 5-15\%$ (stress drop) and $\beta \approx 99\%$ (corresponding
to a seismic efficiency of $1\%$).
The distribution of event sizes also follows a power law (Gutenberg-Richter
for aftershocks), but in contrast to Omori's law, the exponent continuously
depends on the
three parameters. Furthermore, for large dissipation $\beta$, the power law
extends only up to a maximum scale which decreases as $\beta \to 1$.

These results can be rationalized by the following
mean field theory. From (\ref{af:ca-decay-law}), we see that the time
$\Delta t$ needed for an isolated element to reach rupture is such that
the threshold $B\left( \overrightarrow{x},t_0 + \Delta t\right)$
decreases to the stress level $V(\overrightarrow{x})$. The mean field
argument simply assumes that we can extend this result over all elements
of the lattice by replacing $V(\overrightarrow{x})$ by the average stress
$\left< V\lp t\rp  \right>_{\vec{x}}$ accounting for the influence of the
possible loading by n.n. This approximation becomes better and better as
the dissipation $\beta$ increases and the dynamics of n.n. elements becomes
increasingly uncoupled. This corresponds to a decreasing dependence on spatial
inhomogeneities, which is exactly the underlying assumption of any mean-field
approximation. We get
\be
\Delta t \approx B_0^2/\left< V \right>_{\vec{x}}^\alpha~,
\label{qkqkqkk}
\ee
where we have approximated $B_0-\left< V \right>_{\vec{x}}$ by
$B_0$, since for large times the mean stress level becomes very low
compared to the thresholds that are healed back to a typical value in
the interval $B_{0}[1-r,1+r]$ after each event. Expression (\ref{qkqkqkk})
has the
same form as obtained from a model of cracks undergoing
sub-critical crack growth \cite{Matt}.
Over such a time interval, essentially one main event occurs on each site
and, as a
consequence,
 the average stress goes from $\left< V \right>_{\vec{x}}$
to $(1-\gamma)\left< V \right>_{\vec{x}}$, corresponding to a typical
stress decrease $\gamma \left< V \right>$. We can thus write
\be
{d\left< V \right>_{\vec{x}} \over dt} \sim -{\gamma \left< V
\right>_{\vec{x}} \over \Delta t}
 \approx -{\gamma \over B_0^2}~\left< V \right>_{\vec{x}}^{1+\alpha}~,
 \ee
 whose solution is
\be
\left< V \lp t\rp \right>_{\vec{x}} = (B_0^2 /
\alpha\gamma)^{1/\alpha}~(t+c)^{-1/\alpha}~.
\ee
$c$ is a constant determined from the initial value of the average stress
at the beginning of the aftershock relaxation sequence.

To get Omori's law, we recognise that the rate $n(t)$ of aftershocks is simply
proportional to
the rate with which the thresholds $B\left(\overrightarrow{x},t\right)$ reach
the stress level. $n(t)$ is also proportional to $1/\Delta t$.
This yields
$n(t) \propto d B\lp \overrightarrow{x},t\rp / dt \sim
[V\lp \overrightarrow{x},t\rp]^{\alpha}
\sim [\left< V\lp t\rp \right>_{\vec{x}} ]^{\alpha}
\sim {1 \over \lp t+c\rp ^p}$, with $p=1$.
According to this mean field theory, Omori's law
is obtained with the universal exponent $p=1$ independently of the value of the
stress-corrosion exponent $\alpha$, as long as it is positive.
For $\alpha=0$, the average stress level
decays exponentially fast with time and the rate of aftershocks is constant.

The mean field theory also provides a prediction of log-periodicity.
The stress redistribution laws
(\ref{hfql}) and (\ref{hfjqkq}) imply
that, over a typical time $\Delta t$ given by (\ref{qkqkqkk}),
the average stress undergoes the change
$\left< V \right>_{\vec{x}} \to \left< V \right>_{\vec{x}} / \mu$, where
\be
1/\mu = \left[f(d)(1-\gamma) + n_{\rm eff}\gamma(1-\beta) \right]/f(d)~.
\label{jkmqmmq}
\ee
$f(d)$ is a geometric factor counting the effective number of
n.n. ($f(d)=2d$ in the large dissipation limit $\beta \to 1$). The first
contribution
$(1-\gamma)$ in the r.h.s. of (\ref{jkmqmmq}) is simply the initial stress
minus
the stress drop.
The second contribution $n_{\rm eff}\gamma(1-\beta)/f(d)$ results from the
number $n_{\rm eff} \sim 1$ of stress loading on a given element
due to the earthquakes occurring on its neighbours.

Each time the average stress is
decreased by a factor $\mu$, we see from (\ref{qkqkqkk})
that the time interval $\Delta t$ is increased by a factor
\be
\lambda = \mu^{\alpha}~.
\label{mmmqmmqq}
\ee
Since $\mu >1$, the total time is essentially dominated by
the last time interval between the two last cycles. This allows us to write an
approximate scaling relation on the average stress $\left< V \right>$\,:
$\left< V \lp t \rp \right>_{\vec{x}}  = \mu ~\left< V \left( \lambda t\right)
\right>_{\vec{x}}$.
Since the aftershock rate $n(t)$ is proportional to $\left< V
\right>_{\vec{x}}^{\alpha}$ $\mu = \lambda$, this leads to
$n\left( t\right)  = \lambda ~n\left( \lambda t\right)$,
using (\ref{mmmqmmqq}). Looking for a power law solution
$n(t) \sim t^{-p}$, we retrieve Omori's exponent $p=1$. A more general
solution is the power law $t^{-1}$ multiplied by a periodic
function of $\ln t$, $n(t) = t^{-1} P_1(\ln t /\ln \lambda)$, where
$P_1(x)$ is periodic with period one. Expanding
this periodic function into its Fourier series gives
$n\left( t\right) = t^{-1} \sum _{k=-\infty}^{+\infty} a_{k} t^{i2\pi
k/ln\lambda}$,
with $a_{-k} = a_{k}$,
which defines the discrete spectrum of complex exponents
$p_k = 1 - i\frac{2\pi k}{\ln\lambda}$. The
leading correction to the power Omori's law gives the log-periodic expression
$n\left( t\right) =\frac{1}{t} \left( a_{0}+a_{1}\cos
\left( 2\pi \frac{\ln t}{\ln \lambda }\right) \right)$.

\begin{figure}
\begin{center}
\epsfig{file=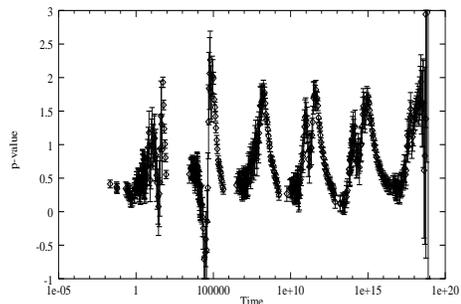, width=6cm, height=4cm}
\caption{\protect\label{movlogp-999} Local exponent $p(t)$ for
$\alpha =1, \gamma=1 , \beta =0.999$. The window size is $100$ events.}
\end{center}
\end{figure}

To test this prediction, we find that the local exponent $p(t)$, defined by
$d \ln n\lp t\rp/d\ln t = - p\lp t \rp$, gives the most sensitive measure
of deviation from the $1/t$ Omori's law. We estimate $p(t)$ by
a maximum likelihood estimator in a running window ending at $t$ \cite{Matt}.
Defining the starting $t$ and ending $t_{U}$ times of
a window and the average $\left< \ln t_{i}\right>$ of the
logarithms of all $N$ aftershock times within this window, the MLE is
$p(t) \approx 12 (\ln\sqrt{t t_U}-\langle \ln t_i \rangle)/
(\ln (t_U/t))^2+1 )$, with a variance
$\sigma^2 \approx 12 \left( N -1\right)^{-1}(\ln (t_{U}/t))^{-2}$.
The estimation of $p(t)$ is very robust
over a large set of window sizes and have been tested thoroughly on synthetic
Omori's laws \cite{methodology}.

Figure \ref{movlogp-999} shows the local exponent $p\lp t\rp$ as a function
of time $t$ estimated using a window size of $100$ events. Clear
log-periodic oscillations around $p\approx 1$ can be identified with a
(log-)frequency of $\approx 0.12$ giving a prefered scale factor
$\lambda \approx 3500$ in reasonable agreement with the theoretical value of
$4000$ calculated from (\ref{jkmqmmq},\ref{mmmqmmqq}). Increasing the window
size with $t$ or keeping a window with size fixed in time provides the same
estimate.

\begin{figure}
\begin{center}
\epsfig{file=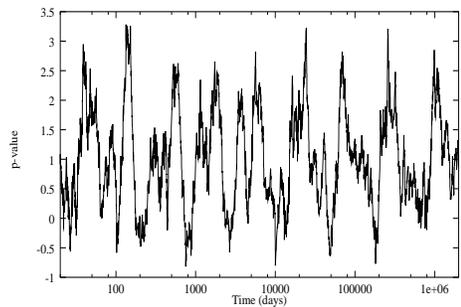, width=6cm, height=4cm}
\caption{\protect\label{smallgamp} $p$-value as a function of time for the
case of \protect\( \alpha =25,\, \gamma =0.05,\, \beta =0.99.\protect \)
The window size is $150$ events.}
\end{center}
\end{figure}

Figure (\ref{smallgamp}) presents $p(t)$ for the stress
corrosion exponent $\alpha \approx 25$, a value estimated from a set
of adjacent time-delayed multiple events in western Japan \cite{das81}, for
a stress drop $\gamma = 5\%$ and a seismic efficiency of $1\%$, i.e. $\beta
=0.99$.
The measured scaling factor is now $\lambda =3.5$ while the mean field
prediction is
$\lambda =3.6$.

\begin{figure}
\begin{center}
\epsfig{file=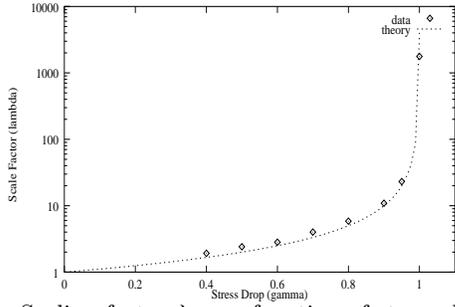, width=6cm, height=4cm}
\caption{\protect\label{logp-datalg}Scaling factor $\lambda$ as a function
of stress
drop $\gamma$ for $\beta =0.999$ and $\alpha =1$.}
\end{center}
\end{figure}

The comparisons between the numerical simulations and the predictions
of the mean field theory are presented in figures 4-6.
The mean field theory, while not perfect, accounts well for most of the
behavior.
We also have checked that  $\lambda$ depends on the space dimension $d$ for
$d=1,2,3$
as predicted from (\ref{jkmqmmq},\ref{mmmqmmqq}).

\begin{figure}
\begin{center}
\epsfig{file=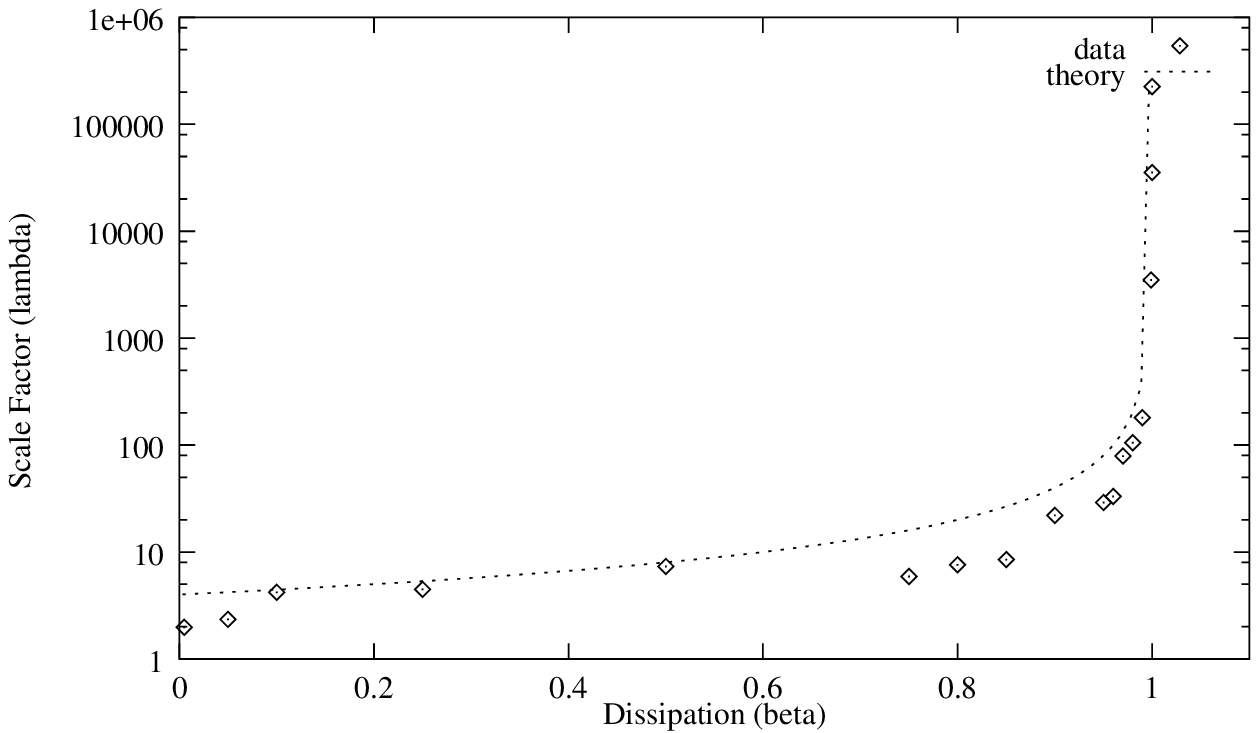, width=6cm, height=4cm}
\caption{\protect\label{lambdbeta} Scaling factor $\lambda$ as a function
of $\beta$ for $\alpha=1$ and $\gamma = 1$.}

\epsfig{file=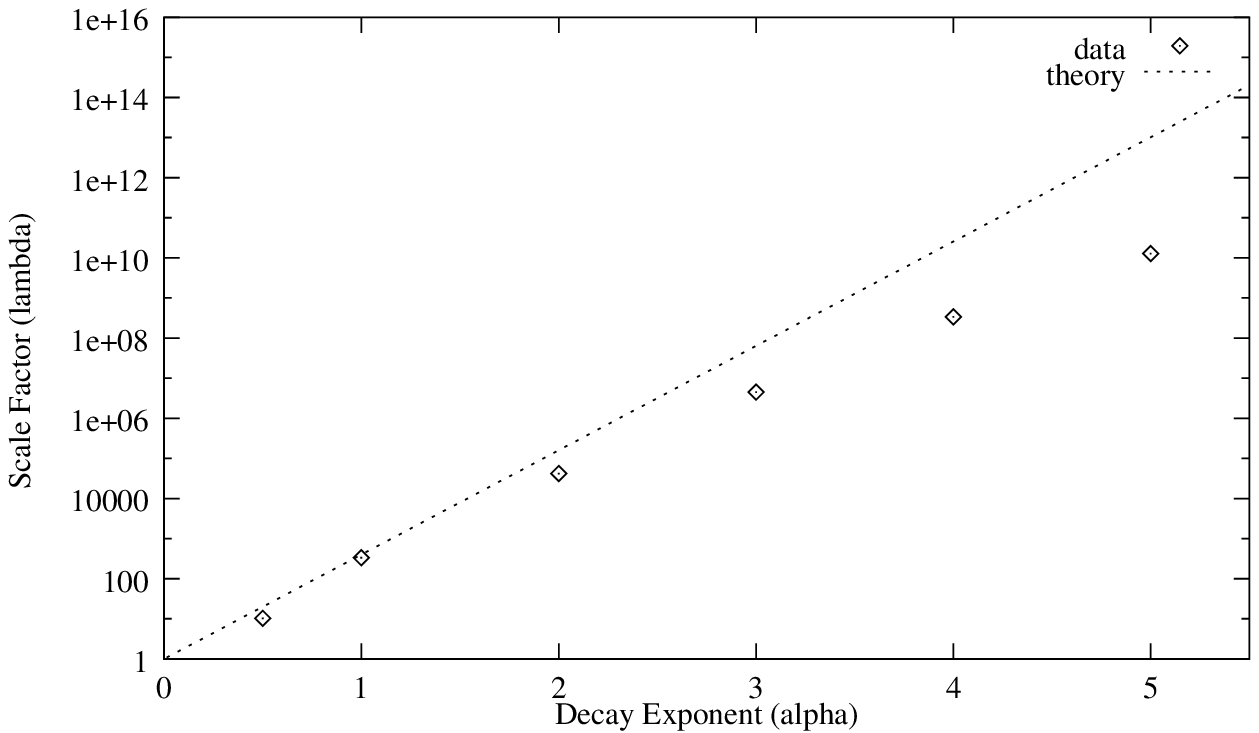, width=6cm, height=4cm}
\caption{\protect\label{lambdalpha} Scaling factor $\lambda$ as a function
of $\alpha$ for $\beta=0.99$ and $\gamma = 1$.}
\end{center}
\end{figure}

Discrete scale invariance
and its log-periodic signature is associated in this model to the threshold
nature of the dynamics. The discrete scaling emerges from the fact that the
thresholds are healed back to a value close to $B_0$ after each rupture and
then have to
decay down to the current stress threshold to trigger the next rupture.
When this
occurs, the stress jumps to a smaller value by a finite amount and can also
latter be reloaded
again by active neighbours. It is fundamentally these {\it finite}
jumps in the stress proportional to the current stress (which are thus
scale-free) which are
at the origin of discrete scale invariance and log-periodicity.
This novel mechanism of log-periodicity does not rely on
a pre-existing discrete structural
hierarchy of faults but is dynamical and reflects the existence of
an approximately fixed stress drop together with the scale-free stress
corrosion power law acting during inter-seismic phases.

This study suggests to search for log-periodic signatures in real
aftershock sequences, with the potential bonus that log-periodicity would
constrain the stress drop ratio, an elusive quantity
to estimate by direct seismic measurements.
A systematic analysis
on more than thirty large aftershock sequences found some indications
\cite{Matt} but
was unable to conclude decisively yet \cite{methodology}, due to
a noise level which is comparable to the amplitude of the signal. Further
studies
are thus called for with better quality data.

We are especially grateful to A. Johansen and L. Knopoff for very
stimulating discussions.
M.L. thanks the Physics department at UCLA for hospitality.

\end{document}